\documentclass[twocolumn,showpacs,preprintnumbers,amsmath,amssymb]{revtex4}

\usepackage{graphicx}

\begin{document}
\title{From Cosmology to Cold Atoms: \\
 Observation of Sakharov Oscillations in Quenched Atomic Superfluids}
\author{Chen-Lung Hung$^{1, *}$}
\author{Victor Gurarie$^2$}
\author{Cheng Chin$^1$}
\address{$^1$ The James Franck Institute and Department of Physics, University of Chicago, Chicago, IL 60637, USA}
\address{$^2$ Department of Physics, CB390, University of Colorado, Boulder CO 80309, USA}

\date{\today}

\begin{abstract}

Sakharov oscillations, conventionally discussed in the context of early universe evolution and the anisotropy of cosmic microwave background radiation, is the manifestation of interfering acoustic waves synchronously generated in an ideal fluid. Here we report the laboratory demonstration of Sakharov oscillations in a quenched atomic superfluid. We quench the sample by Feshbach tuning and monitor the subsequent density fluctuations at different time and length scales by in situ imaging. Sakharov oscillations are identified as the multi-peak structure in the atomic density power spectrum, resembling that of the cosmic microwave background. We also observe Sakharov oscillations in the time domain, from which we extract the energy dispersion of the superfluid, and determine the sonic horizon of the excitations.
\end{abstract}


\maketitle

In modern cosmology, the complex structure of the world we see today can be traced back to the quantum fluctuations in the early universe \cite{Liddle00}. After the inflation, the fluctuations propagate as acoustic pressure waves in the cosmic fluid \cite{PeeblesYu70, Hu2002, Eisenstein08}. The dynamics, first predicted by Andrei Sakharov for a `baryonic' universe \cite{Sakharov65}, manifest in the anisotropy of the cosmic microwave background (CMB) \cite{WMAP} and the large-scale correlations of galaxies \cite{Eisenstein05}. In particular the acoustic waves imprint an intriguing multi-peak structure in the CMB angular power spectrum, called Sakharov oscillations \cite{Sakharov65, ZeldovichNovikov83} or acoustic oscillations, which provide a wealth of indispensable information to infer the density, composition, and even the future evolution of the universe \cite{Hu2002}.

%Evolution of the acoustic waves freezes during the era of recombination when photons and baryons decouple. The remnant signature of acoustic wave dynamics manifests in the anisotropies in the cosmic microwave background (CMB) \cite{Hu2002} and the large-scale correlations function of galaxies \cite{Peebles81, Eisenstein05}. In particular, the acoustic propagation of the cosmic fluid imprints an intriguing multi-peak structure in CMB angular power spectrum, an oscillatory behavior first anticipated by Andrei Sakharov in 1965 \cite{Sakharov65} even for a cold baryonic universe \cite{}. providing a wealth of invaluable information to determine the density, composition and even the destination of the universe \cite{Hu2002}.

Remarkably, the evolution of the early universe depends solely on hydrodynamics and the equation of state, and is insensitive to the microscopic details. Demonstration of Sakharov oscillations in laboratory conditions is possible \cite{Grishchuk12}. In our analogy, the role of gravitational pull and radiation pressure in the cosmic fluid can be captured respectively by the Boson bunching and atomic repulsive interaction in a superfluid \cite{Pethick}. Gravitational instability after inflation can be simulated by a quench of the atomic interaction. In both systems, excitations propagate hydrodynamically as acoustic waves which can superimpose and interfere. These features underlie many intriguing ideas \cite{CosmoAnalogy} and experiments \cite{Lahav2010,Jaskula2012} to associate cosmology and black hole physics to the dynamics of quantum gases.

% Bosonic quantum gas, for instance, can be an excellent candidate to simulate Sakharov oscillations in the early universe: Instability caused by gravitational pull and radiation pressure in a cosmic fluid is simulated by the imbalance between Bosonic stimulation and $s-$wave repulsion in an atomic superfluid.

Sakharov oscillations result from interfering acoustic waves that are synchronously generated throughout a fluid \cite{Hu2002}. The synchronous generation ensures the phase coherence of the acoustic waves \cite{Albrecht96}, while the sound speed $v$ relates the time and length scales of the wave dynamics. Assume that two counter-propagating waves with momenta $\hbar k$ and $-\hbar k$ are created with a relative phase $\phi$, where $2\pi\hbar$ is the Planck constant. After propagating for a time $\tau$, the waves interfere constructively when $2kv \tau+\phi=2m\pi$ and destructively when $2kv\tau+\phi=2(m-1/2)\pi$, where $m=1,2,3..$. Notably, $k_c=\pi/v\tau $ defines the sonic horizon and Sakharov oscillations occur in the ``sub-horizon regime'' $k>k_c$. In the ``super-horizon regime'' $k<k_c$, no interference is expected \cite{Hu2002}.

\begin{figure}[t]
\includegraphics[width=3.4 in]{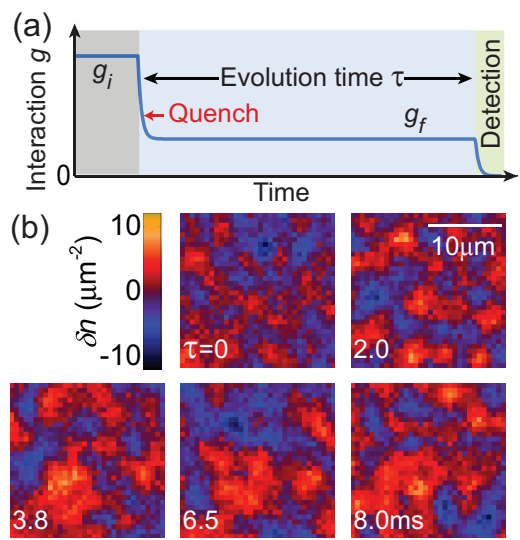}
\caption{(color online). Experimental sequence and density fluctuations of quenched atomic superfluids. (a) the interaction strength $g$ is initially held at $g_i$ and is quenched to a new value $g_f$ by Feshbach tunning. Subsequently, the interaction is held for an evolution time $\tau~$ and the sample is imaged. (b) shows the evolution of the density fluctuation $n-\bar{n}$ of a quenched superfluid with $g_i=0.25$ and $g_f=0.079$. Mean density of the sample is $\bar{n}=$~11~$/\mu$m$^2$.} \label{fig1}
\end{figure}

In this letter, we report the observation of Sakharov oscillations in quenched atomic superfluids. To synchronously generate sound waves in a superfluid, we quench the atomic interactions by Feshbach tuning \cite{Chin2010}. We then monitor the density responses by in situ imaging. The density fluctuations of the sample show a multi-peak structure in the power spectrum, resembling that of the CMB radiation. From the temporal evolution of the density fluctuations, we further determine the dispersion and the sonic horizon of the superfluid. Based on the Bololiugbov theory, we interpret the oscillations as the interference of phonon pairs created by the quench. Remarkably, theoretical studies on the spatial correlations of fast expanding two-dimensional (2D) Bose gases \cite{Imambekov2009}, and of a quenched Bose gas \cite{Natu2012} suggest similar structures.

Our experiment is based on 2D atomic superfluids. Details of the experimental setup are given in Ref.~\cite{Hung2011nature}. In brief, we laser cool and Bose condense cesium atoms in an optical dipole trap, which is then adiabatically deformed into a highly oblate potential with a high harmonic vibrational frequency of $\omega_z= 2 \pi \times 1900~$Hz in the vertical ($z$-) direction and a low frequency of $\omega_r = 2 \pi \times 9~$Hz in the radial ($r$-) direction. The atomic sample forms an almost pure 2D superfluid with typically $2 \times 10^4$ atoms at an equilibrium temperature $T= 10\sim 15$nK. The sample extends over 30~$\mu$m in the radial direction and, in the $z$-direction, occupies the vibrational ground state with a harmonic oscillator length of $l_z = 200~$nm. The interaction strength of the 2D superfluid is characterized by a dimensionless parameter $g= \sqrt{8 \pi} a/ l_z$ \cite{Petrov2000}, where the scattering length $a$ is tunable via a magnetic Feshbach resonance \cite{Chin2010}.

To induce synchronous phonon excitations, we quench the interaction $g$ from an initial value $g_i$ to a final value $g_f$ by switching the magnetic field. The $95\%$ switching time of the field is below $300~\mu$s, fast compared to all relevant time scales in the radial direction. After the quench, we maintain the interaction at $g_f$ for a variable hold time $\tau$, and perform in situ absorption imaging at $g=0$ to record the atomic density distribution \cite{Yefsah2011, Hung2011njp}. The experimental procedure is illustrated in Fig.~\ref{fig1}(a). Throughout this paper, we study dynamics for short evolution times $\tau < 8~$ms, much shorter than the radial vibrational time scale of $2\pi / \omega_r = 110~$ms, and the mean density is effectively a constant. Figure~\ref{fig1}(b) shows the density fluctuations after quenching the interaction from $g_i=0.25$ to $g_f=0.079$. Here, density fluctuations are evaluated pixel-wise using $\delta n_i = n_i -\bar{n}_i$, where $n_i$ is the atomic density measured on the $i$-th pixel of the imaging camera and $\bar{n}_i$ is the mean atomic density we derive after averaging 25 images.

% We analyze quench dynamics using density distributions measured near the center of the 2D superfluids. %Mean local density $\bar{n}_i$ is evaluated using about 50 images recorded under identical experiment conditions.

% Acoustic wave dynamics can be induced in two ways: one by quenching the atomic interaction to a weaker value ($g_f < g_i$), hence reducing the sound speed, and the other by quenching to a stronger value ($g_f > g_i$).
When the interaction is quenched to a smaller value $g_f < g_i$, we observe an apparent growth of density fluctuations, both in amplitude and in its length scale, as hold time increases; see Fig.~\ref{fig1}(b). This trend is consistent with the expectation that the superfluid is evolving toward the weak interaction regime \cite{Hung2011njp}. On the other hand, for quenches to a larger interaction strength $g_f > g_i$, we observe an opposite trend with decreasing amplitude and length scale in the density fluctuations.

\begin{figure}[t]
\includegraphics[width=3.6 in]{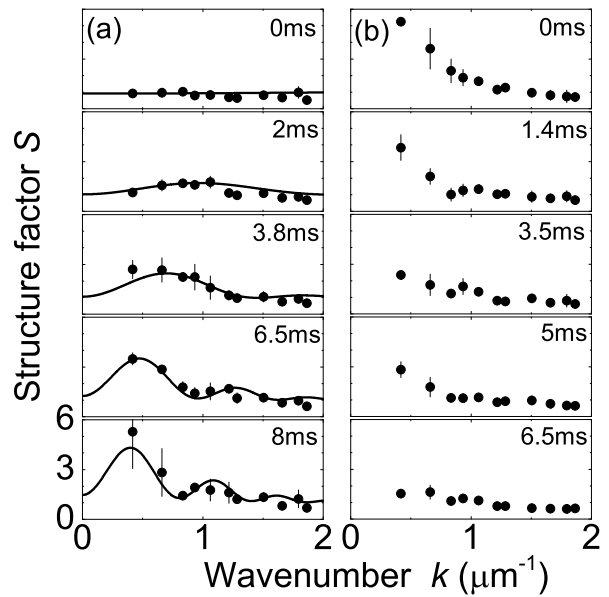}
\caption{Density structure factor of quenched superfluids. The structure factors $S(k)$ (solid circles) are measured at different hold times after quenches from (a) $g_i=0.25$ to $g_f=0.079$ and from (b) $g_i=0.079$ to $g_f=0.14$. In (a), Sakharov oscillations manifest as the multiple-peak structure appearing after 6.5~ms. Solid lines are fit to the experiments; see text. For quench-up measurements, shown in (b), oscillations are less clear. Mean atomic density is $\bar{n}=$11$/\mu$m$^2$ in (a), and 14$/\mu$m$^2$ in (b). }\label{fig2}
%Density fluctuations of quenched superfluids. (a) Quenches from $g_i=0.25$ to $g_f=0.079$: measured structure factors $S(k)$ (solid circles) after different hold times. Sakharov oscillations manifest as the multiple-peak structure appearing after 6.5~ms. Solid lines are fit to the experiments; see text. (b) Quenches from $g_i=0.079$ to $g_f=0.14$. Mean atomic density is $\bar{n}=$11$\mu$m$^{-2}$ in (a), and 14$\mu$m$^{-2}$ in (b). }\label{fig2}
\end{figure}

To study the evolution of the density fluctuations and search for Sakharov oscillations, we evaluate the density structure factor $S(\mathbf{k}) = \frac{\langle |\delta n (\mathbf{k})|^2\rangle}N$, defined as the power spectrum of the density fluctuations \cite{Landau}. Here $\delta n(\mathbf{k})$ is the Fourier transform of the density fluctuation $\delta n(\mathbf{r})$ in real space, $\mathbf{k}$ is the momentum wave vector, and $N$ is the total particle number. The structure factor $S(\mathbf{k})$ is analogous to the angular power spectrum in CMB when a small patch of the sky is analyzed.

% $S(\mathbf{k}) =  \int \nu(\mathbf{r}) e^{-i \bar{k} \cdot \bar{r}} d \mathbf{r}$ is the Fourier transform of the two-point correlation function, and

We evaluate the structure factor based on the central $32\times 32$ pixels of the atomic images, where the mean density $\bar{n}$ is almost uniform. We then perform discrete Fourier transform $\delta n(\mathbf{k}) = A \sum_j \delta n_j e^{- i \mathbf{k} \cdot \mathbf{r}_j}$, where $A = (0.66~\mu$m$)^2$ is the camera pixel size on the object plane. Care has been taken to remove background fluctuations due to photon shot-noise, as well as imaging distortions introduced by optical aberrations \cite{Hung2011njp}. We further bin the spectrum according to the wavenumber $k=|\mathbf{k}|$. Our extraction of $S(k)$ is limited to $k\leq 2 /\mu$m by the imaging resolution and to $k> 0.3 /\mu$m by the resolution of the discrete Fourier transform.

Strong enhancement of the density fluctuations and Sakharov oscillations are observed after we quench the interaction $g$ to a smaller value. Figure~\ref{fig2}(a) shows the density fluctuation spectra after the quench, extracted from images as shown in Fig.~\ref{fig1}(b). At $\tau=0$, the initial power spectrum of the thermal sample at $g_i=0.25$ is rather featureless, as a result of strong interaction and a short correlation length \cite{Hung2011njp}. After the quench, the fluctuations amplify: a peak in the spectrum quickly appears in the first few milliseconds, and its location moves toward smaller $k$ as time $\tau$ increases. This trend suggests that the correlations are spreading out at a finite speed. Spreading of correlations was also reported in a quenched atomic Mott insulator \cite{Cheneau2012}. At longer hold times ($\tau > 5~$ms), the second peak emerges at the detectable range of the power spectrum. The multiple peaks and troughs represent the Sakharov oscillations in the superfluid, resembling those in the angular spectrum of CMB radiation, and support the picture that phonons are created coherently and can interfere at later times.

For quenches toward stronger atomic interactions, shown in Fig.~\ref{fig2}(b), we observe a fast suppression of the fluctuations as the propagation time $\tau$ increases. This trend is consistent with the evolution of the superfluid toward the strong interaction regime. We find a similar time scale for the evolution of the structure factor, but no clear oscillations are observed. Further evidence of Sakharov oscillations will be presented below in the time-dependence of the correlations.

We develop a theoretical understanding of the quench dynamics by the Bogoliubov theory of a weakly interacting Bose gas (supplementary information). At low temperatures, the structure factor can be expressed in the quasi-particle basis as

\begin{eqnarray}
S(\mathbf{k},\tau) &=&\frac{\hbar^2 k^2}{2m \epsilon(\mathbf{k})}  \Big( \langle \hat{b}^\dagger_{\mathbf{k}} \hat{b}_{\mathbf{k}} \rangle + \langle \hat{b}_{-\mathbf{k}} \hat{b}^\dagger_{-\mathbf{k}} \rangle \nonumber\\
&+& e^{2i\epsilon(\mathbf{k}) \tau/\hbar} \langle \hat{b}^\dagger_{\mathbf{k}} \hat{b}^\dagger_{-\mathbf{k}} \rangle + e^{-2i\epsilon(\mathbf{k}) \tau/\hbar } \langle \hat{b}_{-\mathbf{k}} \hat{b}_{\mathbf{k}} \rangle \Big),\label{sc_main}
\end{eqnarray}

\noindent where $\epsilon(\mathbf{k}) = v\hbar k \sqrt{ 1 + k^2\xi^2/4}$ is the Bogoliubov dispersion relation, $\hat{b}_{\mathbf{k}}$ ($\hat{b}_{\mathbf{k}}^\dagger$) is the annihilation (creation) operator of quasi-particles at momentum $\hbar \mathbf{k}$, $v = \frac{\hbar}{m} \sqrt{ng}$ is the speed of sound, $\xi = \hbar/mv$ is the healing length, and $m$ is the atomic mass.

Before the quench, the atomic superfluid is in thermal equilibrium, and the structure factor is given by $S_i(\mathbf{k}) = \frac{\hbar^2\mathbf{k}^2}{2m\epsilon_i(\mathbf{k})}\coth \frac{\epsilon_i(\mathbf{k})}{2k_BT}$ \cite{Landau}, where $\epsilon_i(\mathbf{k})$ is the initial dispersion; see supplementary material. The quench projects the initial state onto a new quasi-particles basis, and, after an evolution time $\tau$, quasi-particles of opposite momenta can interfere, as indicated by the time dependent terms in Eq.~(\ref{sc_main}). The structure factor $S(\mathbf{k}, \tau)$ after the quench is calculated as

\begin{eqnarray}
S(\mathbf{k}, \tau) &=& S_i(\mathbf{k}) \Big[ 1+ \frac{\epsilon_i^2(\mathbf{k})-\epsilon^2(\mathbf{k})}{\epsilon^2(\mathbf{k})} \sin^2 \frac{\epsilon (\mathbf{k}) \tau}{\hbar}\Big]. \label{sd_main}
\end{eqnarray}
%where $\epsilon(\mathbf{k})$ is the dispersion after the quench.
This result suggests a series of acoustic peaks and troughs at $\epsilon(\mathbf{k})\tau/\hbar = \pi/2, \pi...$, which we identify as Sakharov oscillations in atomic quantum gases.

\begin{figure}[h]
\includegraphics[width=3.4 in]{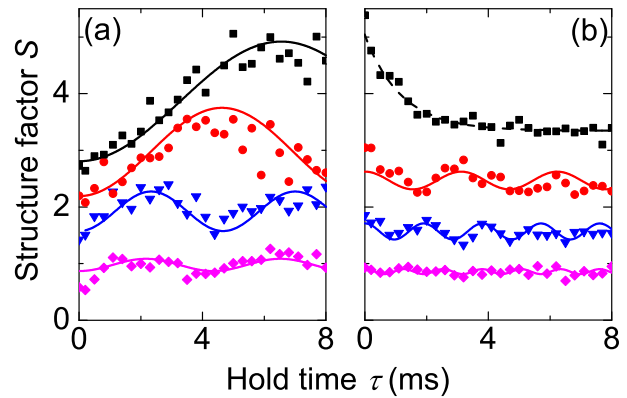}
\caption{(color online). Oscillation of the density structure factor in the time domain. (a) quenches from $g_i=0.25$ to $g_f=0.079$: structure factors $S(k, \tau)$ measured at wavenumbers $k=$0.7 (squares), 0.8 (circles), 1.1 (triangles), and 1.3 (diamonds)~$/\mu$m are shown with an offset of 0.5. (b) quenches from $g_i=0.079$ to $g_f=0.14$: $S(k, \tau)$ measured at $k=$0.7, 0.9, 1.3, 1.6~$/\mu$m. Solid lines are fit to the measurements to determine the oscillation frequencies; see text. The dashed line in (b) is an exponential fit.} \label{fig3}
\end{figure}

The above theoretical form of $S(\mathbf{k},\tau)$ captures well the experimental structure factor after quenches toward weaker interactions, as shown in Fig.~\ref{fig2}(a). We find that the locations of the first and the secondary peaks agree well with the Bogoliubov prediction based on the calculated dispersion and the hold time $\tau$. This agreement supports the picture that counter-propagating quasi-particles coherently interfere after the quench. The observed oscillation amplitudes at larger $k$ are, however, lower than the prediction. The deviations are likely caused by dephasing of phonons at large momenta or residual imaging imperfections \cite{systematic}. For quenches toward larger interaction strengths, the Bogoliubov theory fails to describe our measurement, as will be discussed below.

Sakharov oscillations also manifest in the temporal evolution of the structure factor. Based on the same set of measurements as in Fig.~2, we show the time evolution of $S(k,\tau)$ for various fixed wavenumbers $k$; see Fig.~3. Oscillatory behavior in $S(k,\tau)$ is evident for quenches toward either smaller or larger interaction strengths $g_f$; see Fig.~3 (b). We attribute the ease of observing Sakharov oscillations in the time domain over $k$-space to the higher temporal resolution of our experiment. By and large, we observe as many as 3 oscillations at various $k$ within the 8~ms evolution time. The only exceptions are cases with very small wavenumbers, e.g., $k \leq 0.7~/\mu$m, for which the oscillation periods are expected to be long and the oscillations may be over-damped.

\begin{figure}[h]
\includegraphics[width=3.4 in]{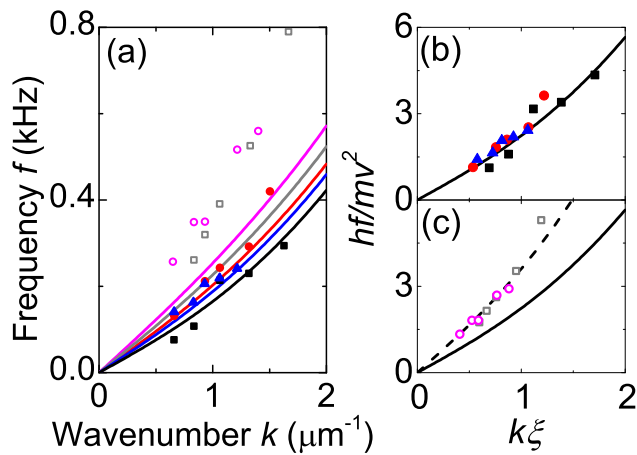}
\caption{(color online). Sakharov oscillation frequencies and energy dispersion. (a) shows the oscillation frequencies determined from $S(k, \tau)$ of quenches to weaker (solid symbols) and stronger (open symbols) interactions. For quenches to smaller interactions, the sample is prepared with $\bar{n}=12/\mu$m$^2$ and $g_i=0.25$, and the final interaction strength is $g_f =$ 0.079 (squares), 0.1 (triangles), and 0.13 (circles). For quenches toward larger interactions, we begin with density $\bar{n} = 13.5~/\mu$m$^2$ and $g_i=0.079$, and quench the interaction to $g_f=$0.14 (open squares) and 0.19 (open circles). (b) and (c) show the scaled dispersion relation based on the measured frequencies.
%Speed of sound $v$ of the sample is 0.45 (squares), 0.55 (triangles), 0.59 (circles), 0.67 (open squares) and 0.76 (open circles)~mm/s.
Solid lines are theory predictions based on the Bogoliubov dispersion $h f/mv^2 = 2 k\xi \sqrt{1 + k^2\xi^2/4}$. Dashed line is an empirical fit which scales up the Bogoliubov result by a factor of 1.65.}
\label{fig4}
\end{figure}

The oscillation frequency of the structure factor in the time domain reflects the energy dispersion of the system, as suggested by Eq.~(\ref{sd_main}). Adopting simple sinusoidal fits, shown as solid lines in Fig.~\ref{fig3}(a) and (b), we determine the oscillation frequencies for various $k$ and $g_f$. In Fig.~\ref{fig4}, we summarize 5 sets of measurements on atomic superfluids quenched to either smaller or larger atomic interactions. We compare our results with Eq.~(\ref{sd_main}), which suggests the oscillation frequency corresponds to twice the phonon energy $f=2\epsilon(\mathbf{k})/h$. Good agreement with the Bogoliubov theory is obtained for quenches toward weaker interactions. This result is fully consistent with the momentum spectra, shown in Fig.~\ref{fig2}(a), where the peak feature in $k$-space also follows the theory well. Measurements for quenches to stronger interactions, however, show significantly higher frequencies than those indicated by the Bogoliubov theory. This discrepancy between quench-down and quench-up can be clearly seen when we further plot the frequency spectra in the scaled units;
 % $hf/mv^2$ and $k\xi$, with $mv^2$ the atomic interaction energy and $\xi$ the healing length
 see Figs.~\ref{fig3}(b) and (c). In the scaled units, our measurements collapse to a single curve, which, for quench-down experiments $g_f<g_i$, is consistent with the Bogoliubov dispersion; for quench-ups $g_f>g_i$, the data overlap, but are about 65$\%$ above the Bogoliubov prediction.

The scaling of the oscillations in the momentum and temporal domains suggests a universal energy-momentum relationship of the excitations in a 2D superfluid and a coherent and self-similar acoustic propagation of the density fluctuations. From the oscillation periods, we confirm the sonic horizon as $k=\pi/v\tau$ for small wavenumber $k$. Our result shows good agreement with the Bogoliubov theory when the interaction is quenched to smaller values. Here, we observe an enhancement and Sakharov oscillations of the density correlations, similar to the expected behavior in the early universe. When the interaction is quenched to larger values, Sakharov oscillation is observed in the time domain. Measured dispersions, however, deviate from the Bogoliubov prediction by $\sim$65$\%$ when the system is left at large interaction strengths. Detailed study of the fluctuations in superfluids with large interactions will be performed and reported elsewhere. 

%Here, the speed of sound $c$ $n\hbar^2g/m=h \times$ 69 (square), 96 (triangle), 116 (circle)~kHz. 146 192~kHz.
%11.4, 12.6, 11.7

We are grateful to Leo Radzihovsky, Wayne Hu and Chao-Lin Kuo for helpful discussions. We thank X. Zhang, L.-C. Ha and S.-K. Tung for the laboratory support. C.-L.~H. and C.~C. are supported by NSF Award PHY-1206095, ARO Grant W911NF0710576 with funds from the DARPA OLE Program, and the Packard foundation. V.G. is supported by  NSF Award PHY-1211914.

$^*$Present address: Norman Bridge Laboratory of Physics 12-33, California Institute of Technology, Pasadena, California 91125, USA.

\section*{Supplementary information}
\subsection{Time dependence of the structure factor $S(\mathbf{k})$}
The density structure factor, defined as the Fourier transform of the density-density correlation function, can be expressed as the correlations in momentum space
\begin{equation}
S(\mathbf{k})= \frac{1}{N} \sum_{\mathbf{q},\mathbf{q}'} \langle \hat{a}^\dagger_{\mathbf{q}+\mathbf{k}} \hat{a}_{\mathbf{q}} \hat{a}^\dagger_{\mathbf{q}'-\mathbf{k}} \hat{a}_{\mathbf{q}'}\rangle, \label{sa}
\end{equation}
where $\hat{a}_\mathbf{k}$($\hat{a}^\dagger_\mathbf{k}$) stands for the annihilation (creation) operator of a momentum state $|\mathbf{k}\rangle$. %The structure factor is then related to the expectation value of first order two-body processes, involving momentum transfer of $\mathbf{k}$, that produce correlated pairs momentum states.
For a degenerate Bose gas, $S(\mathbf{k})$ is dominated by ground state contributions. Substituting $\hat{a}_0$ and $\hat{a}^\dagger_0$ by $\sqrt{N}$, where $N$ is the total atom number, the structure factor reduces to
\begin{equation}
S(\mathbf{k})=  \langle \hat{a}^\dagger_\mathbf{k} \hat{a}_\mathbf{k} \rangle + \langle \hat{a}_{-\mathbf{k}} \hat{a}^\dagger_{-\mathbf{k}} \rangle + \langle \hat{a}^\dagger_\mathbf{k} \hat{a}^\dagger_{-\mathbf{k}} \rangle + \langle \hat{a}_{-\mathbf{k}} \hat{a}_\mathbf{k} \rangle.\label{sb}
\end{equation}
\noindent Here the first two terms relate to the population of $|\pm \mathbf{k}\rangle$ momentum states, and the latter two terms correspond to the correlations between counter-propagating states.

For weakly interacting Bose gases, Eq.~(\ref{sb}) can be further evaluated using the Bogoliubov theory. Here the momentum state operators can be written as the quasi-particle operators $\hat{b}_\mathbf{k}$ ($\hat{b}^\dagger_\mathbf{k}$) under a hyperbolic rotation
\begin{equation}
\begin{pmatrix}
     \hat{a}_\mathbf{k}  \\
     \hat{a}^\dagger_{-\mathbf{k}}
\end{pmatrix}
=
\begin{pmatrix}
     \cosh \alpha_\mathbf{k} & -\sinh \alpha_\mathbf{k} \\
     -\sinh \alpha_\mathbf{k} & \cosh \alpha_\mathbf{k}
\end{pmatrix}
\begin{pmatrix}
     \hat{b}_\mathbf{k}  \\
     \hat{b}^\dagger_{-\mathbf{k}}
\end{pmatrix},\label{rot}
\end{equation}
where $\alpha_\mathbf{k} = \cosh^{-1}\sqrt{\frac{\hbar^2k^2/2m + m v^2}{ 2 \epsilon(\mathbf{k})} + \frac{1}{2}}$ is the rotation angle, $\epsilon(\mathbf{k})=v \hbar k \sqrt{1 + (\frac{\hbar k}{2mv})^2}$ is the energy of the quasi-particle, and $v$ is the sound speed. In the quasi-particle basis, the structure factor reads
\begin{equation}
S(\mathbf{k}) = C_\mathbf{k} \left[\langle \hat{b}^\dagger_\mathbf{k} \hat{b}_\mathbf{k} \rangle + \langle \hat{b}_{-\mathbf{k}} \hat{b}^\dagger_{-\mathbf{k}} \rangle + \langle \hat{b}^\dagger_\mathbf{k} \hat{b}^\dagger_{-\mathbf{k}} \rangle + \langle \hat{b}_{-\mathbf{k}} \hat{b}_\mathbf{k} \rangle \right],
\end{equation}
consisting of similar terms to those in Eq.~(\ref{sb}) except for an overall factor $C_\mathbf{k} = (\cosh \alpha_\mathbf{k} - \sinh \alpha_\mathbf{k})^2 = \hbar^2k^2/2m\epsilon(\mathbf{k})$. Since the operator $\hat{b}^\dagger_\mathbf{k} (t)= \hat{b}^\dagger_{\mathbf{k}} e^{i\epsilon(\mathbf{k}) t/\hbar }$  ($\hat{b}_\mathbf{k} (t)= \hat{b}_{\mathbf{k}} e^{-i\epsilon(\mathbf{k}) t/\hbar }$) is the positive (negative) frequency solution of the equation of motion \cite{Pethick} $\frac{d^2 }{dt^2} \delta \hat{n}= \frac{\epsilon(\mathbf{k})^2}{\hbar^2} \delta \hat{n}$, we can write down the time-dependent form of the structure factor as
\begin{eqnarray}
S(\mathbf{k}) &=&\frac{\hbar^2 k^2}{2m \epsilon(\mathbf{k})}\Big[\langle \hat{b}^\dagger_{\mathbf{k}} \hat{b}_{\mathbf{k}} \rangle + \langle \hat{b}_{-\mathbf{k}} \hat{b}^\dagger_{-\mathbf{k}} \rangle \nonumber\\
 &+& e^{2i\epsilon(\mathbf{k}) t/\hbar} \langle \hat{b}^\dagger_{\mathbf{k}} \hat{b}^\dagger_{-\mathbf{k}} \rangle + e^{-2i\epsilon(\mathbf{k}) t/\hbar } \langle \hat{b}_{-\mathbf{k}} \hat{b}_{\mathbf{k}} \rangle \Big].\label{sc}
\end{eqnarray}
From Eq.~(\ref{sc}), the structure factor evolves with time only when there is correlation between counter-propagating quasi-particle pairs.

\subsection{Structure factor at thermal equilibrium}

At thermal equilibrium, the number of quasi-particles obeys Bose-Einstein statistics
\begin{eqnarray}
\langle \hat{b}^\dagger_{\mathbf{k}} \hat{b}_{\mathbf{k}} \rangle = \langle \hat{b}_{-\mathbf{k}} \hat{b}^\dagger_{-\mathbf{k}} \rangle -1 =  \frac{1}{e^{\epsilon(\mathbf{k})/k_BT}-1}.\label{BEnum}
\end{eqnarray}
In addition, there is no net source or sink to generate correlated quasi-particles
\begin{equation}
\langle \hat{b}^\dagger_{\mathbf{k}} \hat{b}^\dagger_{-\mathbf{k}} \rangle = \langle \hat{b}_{-\mathbf{k}} \hat{b}_{\mathbf{k}} \rangle=0.\label{BEsource}
\end{equation}
Using Eqns.~(\ref{sc}), (\ref{BEnum}), and (\ref{BEsource}), we find the equilibrium static structure factor \cite{Landau}
\begin{equation}
S(\mathbf{k}) =\frac{\hbar^2 k^2}{2 m \epsilon(\mathbf{k}) } \coth \frac{\epsilon(\mathbf{k})}{2 k_B T}.
\end{equation}

\subsection{Evolution of the structure factor after a quench}

When the interaction is quenched, quasi-particles are projected out from the condensate, causing the structure factor to evolve with time. We find the expectation values in Eq.~(\ref{sc}), expressed in terms of the quasi-particle operator $\hat{c}_{\mathbf{k}}$ ($\hat{c}^\dagger_{\mathbf{k}}$) right before the quench,
\begin{eqnarray}
\langle \hat{b}^\dagger_{\mathbf{k}} \hat{b}_{\mathbf{k}} \rangle = \cosh^2 \Delta \alpha_\mathbf{k} \langle \hat{c}^\dagger_{\mathbf{k}} \hat{c}_{\mathbf{k}} \rangle + \sinh^2 \Delta \alpha_\mathbf{k} \langle \hat{c}_{-\mathbf{k}} \hat{c}^\dagger_{-\mathbf{k}} \rangle, \label{quenchproj1}
\end{eqnarray}
and
\begin{eqnarray}
\langle \hat{b}^\dagger_{\mathbf{k}} \hat{b}^\dagger_{-\mathbf{k}} \rangle = \frac{1}{2} \sinh 2\Delta \alpha_\mathbf{k} \left( \langle \hat{c}^\dagger_{\mathbf{k}} \hat{c}_{\mathbf{k}} \rangle + \langle \hat{c}_{-\mathbf{k}} \hat{c}^\dagger_{-\mathbf{k}} \rangle \right).\label{quenchproj2}
\end{eqnarray}
Similar expressions hold for $\langle \hat{b}_{-\mathbf{k}} \hat{b}^\dagger_{-\mathbf{k}} \rangle$ and $\langle \hat{b}_{-\mathbf{k}} \hat{b}_{\mathbf{k}} \rangle$. Here, $\Delta \alpha_\mathbf{k} = \frac{1}{2} \cosh^{-1} \frac{1}{2} \left[ \frac{\epsilon(\mathbf{k})}{\epsilon_i(\mathbf{k})} + \frac{\epsilon_i(\mathbf{k})}{\epsilon(\mathbf{k})} \right]$ is the hyperbolic angle separation between two bases, and $\epsilon_i(\mathbf{k})$ is the Bogoliubov energy before the quench. Using Eqns.~(\ref{quenchproj1}) and (\ref{quenchproj2}), and applying equilibrium Bose statistics to the initial state population, Eq.~(\ref{sc}) can be written as
\begin{eqnarray}
S(\mathbf{k}) =  &&\frac{\hbar^2 k^2}{2m \epsilon(\mathbf{k})} \coth \frac{\epsilon_i(\mathbf{k})}{2k_B T} \times \nonumber \\
 &&\Big[ \cosh 2 \Delta \alpha_\mathbf{k} + \sinh 2 \Delta\alpha_\mathbf{k} \cos \frac{2 \epsilon(\mathbf{k}) t}{\hbar}\Big],
\label{sd}
\end{eqnarray}
or, equivalently,
\begin{eqnarray}
S(\mathbf{k}) &=&S_i(\mathbf{k}) \Big[ 1+ \frac{\epsilon_i^2(k)-\epsilon(\mathbf{k})^2}{\epsilon(\mathbf{k})^2} \sin^2 \frac{\epsilon(\mathbf{k})t}{\hbar} \Big], \label{sin2}
\end{eqnarray}
where $S_i(\mathbf{k}) = \frac{\hbar^2 k^2}{2m \epsilon_i(\mathbf{k})} \coth \frac{\epsilon_i(\mathbf{k})}{2k_B T}$ is the initial equilibrium structure factor.

Applicability of Eq.~(\ref{sin2}) can also be verified using the density-phase formalism \cite{Mora2003}. This formalism also shows the breakdown of Eq.~(\ref{sin2}) in case when $g_f=0$ and the initial temperature is close to the Berezinsky-Kosterlitz-Thouless superfluid transition, in agreement with Ref.~\cite{Imambekov2009}.


\begin{thebibliography}{99}
\bibitem{systematic} Our fit is based on Eq.~(\ref{sd_main}) with the temperature $T$ as a fitting parameter. In addition, we adopt an empirical suppression factor $G(k)=\exp(-k^2/k_0^2)$ to fit the correlation part of the measured $S_{exp}(k)$, namely, $S_{exp}(k)=1+G(k)(S(k)-1)$. The best fit gives $k_0=1.4/\mu$m.
\end{thebibliography}
\end{document}